\input harvmac
\input psfig
\newcount\figno
\figno=0
\def\fig#1#2#3{
\par\begingroup\parindent=0pt\leftskip=1cm\rightskip=1cm\parindent=0pt
\global\advance\figno by 1
\midinsert
\epsfxsize=#3
\centerline{\epsfbox{#2}}
\vskip 12pt
{\bf Fig. \the\figno:} #1\par
\endinsert\endgroup\par
}
\def\figlabel#1{\xdef#1{\the\figno}}
\def\encadremath#1{\vbox{\hrule\hbox{\vrule\kern8pt\vbox{\kern8pt
\hbox{$\displaystyle #1$}\kern8pt}
\kern8pt\vrule}\hrule}}
\def\underarrow#1{\vbox{\ialign{##\crcr$\hfil\displaystyle
 {#1}\hfil$\crcr\noalign{\kern1pt\nointerlineskip}$\longrightarrow$\crcr}}}
%
\overfullrule=0pt
\def\hat{\widehat}
%
\def\tilde{\widetilde}
\def\bar{\overline}
\def\Z{{\bf Z}}
\def\T{{\bf T}}
\def\C{{\bf C}}
\def\S{{\bf S}}
\def\R{{\bf R}}

\font\zfont = cmss10 
\font\litfont = cmr6

\def\bigone{\hbox{1\kern -.23em {\rm l}}}
\def\ZZ{\hbox{\zfont Z\kern-.4emZ}}
\def\half{{\litfont {1 \over 2}}}

\Title{hep-th/9912086}
{\vbox{\centerline{Duality Relations}
\bigskip
\centerline{ Among Topological Effects In String Theory}}}
\smallskip
\centerline{Edward Witten $^*$}
\smallskip
\centerline{\it Physics Department, California Institute of Technology,
Pasadena CA 91125 USA}
\smallskip
\centerline{and}
\smallskip
\centerline{\it CIT-USC Center for Theoretical Physics,
Univ. of Southern California, Los Angeles CA}\bigskip
\vskip 2 cm
\noindent
We explore two different problems in string theory in which duality relates an ordinary $p$-form field in one theory to a self-dual $(p+1)$-form field in 
another theory.  One problem involves comparing $D4$-branes to $M5$-branes, and the other involves comparing the Ramond-Ramond forms in Type IIA and Type IIB superstring theory.  In each case, a subtle topological effect involving the $p$-form can be recovered from a careful analysis of the quantum
mechanics of the self-dual $(p+1)$-form. 
 \vskip 3cm
\noindent {$^*$ On leave from Institute for Advanced Study, Princeton NJ 08540
}
\medskip

\Date{December, 1999}
\newsec{Introduction}

The purpose of this paper is to explore how certain relatively
subtle topological effects in string theory and $M$-theory
transform into each other under dualities.   We will look at two
cases that are rather similar and can be treated in rough parallel:

\def\spinc{{\rm Spin}^c}
(1) The ``$U(1)$ gauge field'' on the world-volume of a Type II $D$-brane
is actually better described as a ${\rm Spin}^c$ structure (assuming,
as we generally will in the present paper, that the background Neveu-Schwarz
three-form field $H$ is topologically trivial). 
This effect, which first showed up in a detailed example
\ref\branecft{E. Witten, ``Baryons and Branes In Anti de Sitter Space,''
JHEP 9807:006 (1998), hep-th/9805112.},
has a natural interpretation in $K$-theory
\nref\moore{R. Minasian and G. Moore, ``$K$ Theory And Ramond-Ramond
Charge,'' JHEP 9711:002 (1997), hep-th/9710230.}%
\nref\kwitten{E. Witten, ``$D$-Branes And $K$ Theory,'' JHEP 9812:019
(1998), hep-th/9810188.}%
\refs{\moore,\kwitten}
and can be demonstrated by studying global anomalies for elementary
strings ending on the $D$-brane \ref\freed{D. Freed and E. Witten,  ``Anomalies
In String Theory With $D$-Branes,'' hep-th/9907189.}.
The effect exists for Type IIA and IIB $Dp$-branes for several
values of $p$. The problem we will study arises in the case of a
Type IIA $D4$-brane.  Such a brane can arise
upon compactifying an $M5$-brane on a circle, in which case the
``gauge field'' of the $D4$-brane arises by compactifying the
chiral two-form (with self-dual curvature) on the $M5$-brane.
It must somehow be possible to deduce the
$\spinc$ nature of the $D4$ gauge field 
from some property of the chiral two-form of the $M5$-brane.

(2) The Ramond-Ramond four-form field strength $G_4$ of Type IIA
superstring theory does not, in general, obey conventional Dirac 
quantization.  Under certain conditions \ref\ugwitten{E. Witten,
``On Flux Quantization In $M$ Theory And The Effective Action,''
J. Geom. Phys. {\bf 22} (1997) 1, hep-th/9609122.}, there is a gravitational correction to the
quantization law, and the periods of $G_4$ are half-integral.
Type IIA superstring theory on a spacetime $X=\S^1\times Y$  is
$T$-dual to Type IIB on the same spacetime.
The $T$-duality maps the relevant part of $G_4$ to the
self-dual five-form $G_5$ of Type IIB on $ \S^1\times Y$.
Hence, in this situation,
it must be possible to deduce the nonintegrality of the $G_4$
periods from some property of the dynamics
of $G_5$.

\nref\xxx{N. Marcus and J. H. Schwarz, ``Field Theories That Have No
Manifestly Lorentz Invariant Formulation,'' Phys. Lett. {\bf B115}
(1982) 111.}%
\nref\seigel{W. Siegel, ``Manifest Lorentz Invariance Sometimes
Requires Nonlinearity,'' Nucl. Phys. {\bf B238} (184) 307.}%
 \nref\godol{P. Goddard and
D. Olive, ``Algebras, Lattices, and Strings'' (preprint, 1983),
Phys. Scripta {\bf T15} (1987).}%
\nref\imb{C. Imbimbo and A. Schwimmer, ``The Lagrangian Formulation Of
Chiral Scalars,'' Phys. Lett. {\bf B193} (1987) 455.}%
\nref\hull{C. M. Hull, ``Covariant Quantization Of Chiral Bosons And
Anomaly Cancellation,'' Phys. Lett. {\bf B206} (1988) 234.}%
\nref\laba{J. M. F. Labastida and M. Pernici, ``Lagrangians For Chiral Bosons
And The Heterotic String,'' Nucl. Phys. {\bf B306} (1988) 516.}%
\nref\mez{L. Mezincescu and R. I. Nepomechie, ``Critical Dimensions For Chiral
Bosons,'' Phys. Rev. {\bf D37} (1988) 3067.}%
\nref\hentei{M. Henneaux and C. Teitelboim, ``Dynamics Of Chiral
(Selfdual) $p$ Forms,'' Phys. Lett. {\bf B206} (1988) 650; F. P. Devecchi and M.
Henneaux, ``Covariant Path Integral For Chiral $p$-Forms,'' Phys. Rev.
{\bf D54} (1996) 1606.}
\nref\sriv{P. P. Srivastava, ``On A Gauge Theory Of Selfdual Field And
Its Quantization,'' Phys. Lett. {\bf B234} (1990) 93.}%
\nref\mcclain{B. McClain, Y. S. Wu, and F. Yu, ``Covariant Quantization
Of Chiral Bosons and $OSp(1,1|2)$ Symmetry, Nucl. Phys. {\bf B343} (1990)
689.}%
\nref\wota{C. Wotzase, ``The Wess-Zumino Term For Chiral Bosons,''
Phys. Rev. Lett. {\bf 66} (1991) 129.}%
\nref\mre{I. Martin and A. Restuccia, ``Duality Symmetric Actions And
Canonical Quantization,'' Phys. Lett. {\bf B323} (1994) 311.}
\nref\schwsen{J. H. Schwarz and A. Sen, ``Duality Symmetric Actions,''
Nucl. Phys. {\bf B411} (1994) 35, hep-th/9304154.}%
\nref\everl{E. Verlinde, ``Global Aspects Of Electric-Magnetic Duality,''
Nucl. Phys. {\bf B455} (1995) 211, hep-th/9506011.}%
\nref\persch{M. Perry and J. H. Schwarz, ``Interacting Chiral Gauge Fields
In Six Dimensional Born-Infeld Theory,'' hep-th/9611065.}%
\nref\schwao{J. H. Schwarz, ``Coupling A Self-Dual Tensor To Gravity
In Six Dimensions,'' hep-th/9701008.}%
\nref\berk{N. Berkovits, ``Local Actions With Electric and Magnetic
Sources,'' hep-th/9610134, ``Super-Maxwell Actions With 
Manifest Duality,'' hep-th/9612174.}%
\nref\benkl{I. Bengtsson and A. Kleppe, ``On Chiral $p$-Forms,'' 
hep-th/9609102; I. Bengtsson, ``Manifest Duality In Born-Infeld Theory,''
hep-th/9612174.}%
\nref\xxxx{P. Pasti, D. Sorokin, and M. Tonin, Phys. Lett. {\bf B352} (1995) 59,
Phys. Rev. {\bf D52} (1995) R4277, ``On Lorentz Invariant Actions
For Chiral $P$-Forms,'' hep-th/9611100, ``Covariant Action For A $D=11$
Five-Brane With The Chiral Field,'' Phys. Lett. {\bf B398} (1997) 41;
I. Bandos, K. Lechner, A. Nurmagambetov, P. Pasti, D. Sorokin,
``Covariant Action For The Super-Five-Brane Of $M$-Theory,'' hep-th/9701149;
G. Dall'Agata, K. Lechner, and M. Tonin, ``Action For IIB SUpergravity
In Ten Dimensions,'' hep-th/9812170.}%
\nref\yyy{M. Aganagic, J. Park, C. Popescu, and J. Schwarz,
``Worldvolume Action For The $M$-Theory Fivebrane,'' Nucl. Phys. {\bf B496}
(1997) 191, hep-th/9701166.}%
\nref\west{P. S. Howe, E. Sezgin, and P. C. West, ``Covariant Field
Equations Of The $M$ Theory Five-Brane,'' Phys. Lett. {\bf B399} (1997) 49,
hep-th/9702008,  ``The Six-Dimensional Self-Dual Tensor,'' Phys. Lett.
{\bf B400} (1997) 255, hep-th/9702111.}%
\nref\miao{Y.-G. Miao, J.-G. Zhou, and Y.-Y. Liu, ``New Way Of The Derivation
Of First Order Wess-Zumino Terms,'' 
Phys. Lett. {\bf B323} (1994) 169; Y.-G. Miao and H. J. W. Muller-Kirsten,
``Self-Duality Of Various Chiral Boson Actions,'' hep-th/9912066.}%
\nref\maz{A. Mazyntsia, C. R. Preitschopf, and D. Sorokin, ``Dual Actions
For Chiral Bosons,'' hep-th/9808049.}%
\nref\heno{X. Bekaert, M. Henneaux, and A. Sevrin, ``Deformations
Of Chiral Two-Forms In Six Dimensions,'' hep-th/9909094, ``Symmetry-deforming
Interactions Of Chiral $p$-Forms,'' hep-th/9912077.}%
What these examples have in common is that on one side of the
relation, one considers a field (the ``gauge field strength'' on the $D4$-brane,
or the four-form of Type IIA) whose periods are shifted from
conventional Dirac quantization by a gravitational correction.
On the other side of the relation
is a self-dual Bose field of one degree higher
(the three-form of the $M5$-brane, and the five-form of Type IIB)
in a related theory.  We must somehow deduce the gravitational
correction in the lower dimension from the quantum mechanics of the self-dual
field in the higher dimension.

The quantum mechanics of a self-dual field
is quite subtle and has been studied from many points
of view, a sampling being
 \refs{\xxx-\heno}.  Recent work has included construction of brane Lagrangians
at least locally \refs{\persch, \xxxx,\yyy} and
construction of
manifestly supersymmetric and kappa-symmetric equations of motion
for multiplets including the self-dual fields \west.

As is most familiar from the case of a chiral
scalar (self-dual one-form) in two dimensions, and as we will review
in section 3, a chiral $p$-form field generally has
on a given manifold
several possible partition functions, determined by a choice of
theta function; one needs a recipe to pick out the right theta
function in a given situation.
For $p>1$, this has been demonstrated most explicitly in
\ref\hns{M. Henningson, B. E. W. Nilsson, and P. Salomonson,
``Holomorphic Factorization Of Correlation Functions In
$(4k+2)$-Dimensional $(2k)$-Form Gauge Theory,'' hep-th/9908107.}.
The right recipe for picking a theta function depends on
some physical input; for the self-dual three-form of the $M5$-brane,
and the self-dual five-form of Type IIB, a prescription has been
given in \ref\witten{E. Witten, ``Fivebrane Effective Action In
$M$-Theory,'' J. Geom. Phys. {\bf 22} (1997) 103, hep-th/9610234.}.  
For one specific example above two dimensions
--the self-dual three-form on $\T^6$, where the partition function
turns out to be unique (independent of the spin structure on $\T^6$)  -- 
the appropriate partition function has been constructed and
studied in detail \ref\dolanappi{L. Dolan and C. R. Nappi, ``A Modular
Invariant Partition Function For The Five-Brane,'' Nucl. Phys.
{\bf B530} (1998) 683, hep-th/9806016.}.  
The recipe of \witten\ for picking a theta function has been
related to a more classical topological invariant (the Kervaire invariant)
in \ref\hopsinger{M.
Hopkins and I. M. Singer, to appear.}. 

An exception to the statement that
the chiral $p$-form has several possible
partition functions arises \godol\ if one combines several chiral bosons
using an even unimodular lattice.  Then one gets complete modular
invariance and a unique partition function.  This case is
very important for the heterotic string \ref\ghmr{D. J. Gross,
J. A. Harvey, E. Martinec, and R. Rohm, ``The Heterotic String,'' Phys. Rev.
Lett. {\bf 54} (1985) 502.}.  In a different case (like a single chiral scalar
at the free fermion radius, relevant to the present paper),
one cannot resolve the ambiguity of the partition function by
summing  over all possibilities
because each candidate partition function has slightly different anomalies,
and it does not make sense to add them. In the $M$-theory and Type IIB
applications, the chiral $p$-form does not appear by itself but
together with addition fields such as fermions.  The complete partition
function is presumably anomaly-free (this has not been completely
demonstrated); anomaly cancellation 
depends on pairing the proper (spin-structure
dependent) partition function of the fermions with the proper
partition function of the chiral $p$-form.  Thus, one must expect
that the recipe  for picking a chiral $p$-form partition function depends
on the spin structure, and this is the case for the proposal in \witten.
Once the anomalies are all canceled, it is possible, and perhaps correct
physically, to sum over spin structures.

The main goal of the present paper is to show how
the quantum mechanics of the self-dual fields gives rise,
after compactification on a circle, to the effects
mentioned in (1) and (2) above. In section 2,
we demonstrate the phenomena in special cases in which detailed general
theory is not needed.  In the rest of the paper, we proceed
more systematically. In section 3, we recall some important
facts about $p$-form quantum mechanics.  In sections 4 and 5,
we make the theory in \witten\ more concrete for the situation
of interest and use it to deduce what we need.

The first of our two problems described above
is somewhat reminiscent of the problem
of relating the mechanism of $M5$-brane normal bundle anomaly
cancellation \ref\harvey{D. Freed, J. A. Harvey, R. Minasian,
and G. Moore, ``Gravitational Anomaly Cancellation For $M$ Theory
Five-Branes,'' Adv. Theor. Math. Phys. {\bf 2} (1998) 601, hep-th/9803205.}
with the corresponding mechanism in Type IIA \witten.
The relation between them has been analyzed recently
\ref\becker{K. Becker and M. Becker, ``Five-Brane Gravitational
Anomalies,'' hep-th/9911138.}.

\newsec{Reduction To Chiral Scalar}

The goal in the present section is to verify that the phenomena
mentioned in the introduction work out correctly in some simple cases
in which we can do this without many technicalities.  This will
perhaps satisfy the curiosity of some readers, and may give others the
courage needed to persevere through the technicalities of the rest of the paper.

\bigskip\noindent{\it $M5$-Brane Wrapped On A Circle}

We first consider the relation of the $M5$-brane to the $D4$-brane.
Our goal is to analyze
the M5-brane  on a world-volume $V=S\times R$,
where $R$ is an oriented
 five-manifold and $S$ is a circle with a supersymmetry-preserving
spin structure.  To do this in general will be the goal of section 5, 
but things are much simpler in the case $R=\tilde S\times R'$,
with $\tilde S$ another circle and $R'$ a four-manifold.  The simplicity will
arise because in this special case, we do not need to understand
chiral $p$-forms fields of $p>0$; we can deduce what we need from
familiar (though subtle) facts about chiral scalars.

\def\CP{{\bf CP}}
Though we could treat an arbitrary $R'$, it will suffice for
illustration to take $R'=\CP^2$.  Thus, the fivebrane world-volume
will be $V=\Sigma\times \CP^2$ where $\Sigma=S\times \tilde S$ is a product of circles; the spin structure on $S$ preserves
supersymmetry but either choice may be made on $\tilde S$.
The nontrivial cohomology group of $\CP^2$ 
(apart from dimensions zero and four) is
\eqn\togglep{\eqalign{
                      H^2(\CP^2;\Z) & = \Z .\cr}}
The generator of $H^2(\CP^2;\Z)$ is a self-dual form $\omega$ that obeys
\eqn\jukko{\int_{\CP^1}\omega=1,~~\int_{{\bf CP}^2} \omega\wedge \omega = 1.}
Here $\CP^1$ is a linearly embedded subspace of $\CP^2$ and generates
$H_2(\CP^2;\Z)$.

We suppose that the $M$-theory spacetime is $X=\Sigma\times C$,
where $C$ is a nine-dimensional spin-manifold in which $\CP^2$ is embedded.
$M$-theory on this spacetime is equivalent to Type IIA on
$X'=\tilde S\times C$; the M5-brane corresponds to a D4-brane wrapped
on $R=\tilde S\times \CP^2$.  $R$ is not a spin manifold, since $\CP^2$
is not.  As a result, according to \freed, the field strength $F$ of the
``$U(1)$ gauge field'' on the D4-brane does not obey conventional
Dirac quantization.  Rather,
\eqn\pxo{\int_{\CP^1}{F\over 2\pi}= n+\half,}
with integer $n$.  

The gauge field on the D4-brane arises by dimensional reduction from the
chiral two-form $b$ on the M5-brane.  We want to know how \pxo\ arises
from the theory of a chiral two-form.  We consider a limit in which
the radii of $S$ and $\tilde S$ are much greater than the size of the
$\CP^2$.  In this case, the physics on the M5-brane reduces to an effective
two-dimensional theory on $\Sigma=S\times \tilde S$.  In fact, the field
$b$ reduces (by the ansatz $b=\omega \phi$) to a chiral scalar $\phi$
in two dimensions.
$\phi$ appears at the self-dual or free fermion radius\foot{In general,
if the $M5$-brane is compactified to two dimensions on a four-manifold
$R'$, the chiral two-form reduces to a set of two-dimensional scalars
with momentum lattice given by the two-dimensional cohomology
lattice of $R'$.  For $R'=\T^4$, this assertion is built into the detailed
computation in \dolanappi.
For $R'=\CP^2$, the lattice is one-dimensional,
generated by a vector $\omega$ with $\omega^2=1$; this is the lattice
of a chiral boson with the free-fermion radius.  (Depending on how
$R'$ is embedded in the full spacetime, 
some of the conservation laws associated with
the momentum lattice may be violated by instantons constructed from
membranes with boundary on $R'$. 
This phenomenon is irrelevant for determining the fivebrane
partition function in the large volume limit.)}; the $\phi$ field
is hence 
equivalent quantum mechanically to a complex  fermion $\psi$ of positive
chirality.

The $\psi$ field propagates on the Riemann surface $\Sigma$, and
the partition function of $\psi$ depends on a choice of spin structure on
$\Sigma$.
So to describe the physics, we need to know the effective spin structure
on $\Sigma$ in the low energy theory, given the underlying choice
of spin structure on the $M$-theory spacetime $X=\Sigma\times C$.
Since choosing a spin structure on $X$ is equivalent to choosing
a spin structure on $\Sigma$ and choosing one on $C$, in the microscopic
$M$-theory description a spin structure was chosen on $\Sigma$ at the
beginning.  In fact, as we noted above, we are interested in the case
that this spin structure is
the product of the supersymmetric spin structure on $S$ and any
desired spin structure on $\tilde S$.  
It is natural to guess that the effective
spin structure on $\Sigma$ in the low energy theory is just the spin
structure on $\Sigma$ that we start with microscopically.  This
assertion almost follows just from the fact that the map
from the microscopic to the macroscopic spin structure must be invariant
under the action of $SL(2,\Z)$ on $\Sigma$, and can be verified
using the techniques of sections 4 and 5.

In the theory of a $D4$-brane on $\tilde S\times {\bf CP}^2$, with $\tilde S$ regarded
as the ``time'' direction, the flux \pxo\ can be interpreted as
a conserved charge.  Going back to the self-dual three-form theory
on the $M5$-brane worldvolume
$V=S\times \tilde S\times {\bf CP}^2$, this flux is interpreted as the
integral of the self-dual three-form $T$ (which is the curvature of the
chiral two-form $b$, defined by $T=db$) over $S\times {\bf CP}^1$.
In terms of the ansatz $b=\omega \phi$, we have $T=\omega \wedge d\phi$,
and the conserved charge is
\eqn\tikko{q=\int_{S\times \CP^2}{\omega \wedge d\phi\over 2\pi}
 =\oint_S{d\phi\over 2\pi}.}
In the free fermion description, $d\phi/2\pi$ becomes $\bar\psi\psi$
and the charge is the conserved fermion number
\eqn\ulikko{q=\oint_S{\bar\psi\psi}.}
Now, since the fermions on $S$ are in the supersymmetric spin structure,
both $\psi$ and $\bar\psi$ have a single zero mode on $S$.  The
quantization of the zero modes gives rise, in a way that is familiar
from the Ramond sector of superstrings, to a two-fold degeneracy
of the ground state.  The ground states have fermion number $q=\pm 1/2$,
and all excited states have half-integral eigenvalues of $q$.
Since $q$ is interpreted in the Type IIA description as the flux
in \pxo, we have explained the half-integrality of that flux starting
with the theory of the self-dual three-form on the $M5$-brane.

It is also instructive to consider, in a similar fashion, a case in
which the $D4$-brane is wrapped on a five-manifold $R$ that does not have
a $\spinc$ structure, so that the theory should be inconsistent.  Such
a case is obtained by taking $R$ to be not a product $\tilde S\times \CP^2$
but a $\CP^2$ bundle over $\tilde S$ in which the fiber undergoes complex
conjugation in going around $\tilde S$.  Complex conjugation reverses the
sign of $\omega$ and so acts on $\phi$ by $\phi\to -\phi$.  The periods
of $\phi$ thus must change sign in going around $\tilde S$, but since they
are half-integral, this is impossible.  This is the inconsistency.  But
what does it look like in the free fermion description?  From this point
of view, $\phi\to -\phi$ is $\psi\leftrightarrow \bar\psi$.  Alternatively,
if $\psi=(\psi_1+i\psi_2)/\sqrt 2$ with Majorana-Weyl fermions $\psi_1$,
$\psi_2$, it is \eqn\ilmok{\psi_1\to \psi_1,~\psi_2\to -\psi_2.}  
Both $\psi_1$ and $\psi_2$ couple to the supersymmetric spin structure
on $S$, and in view of \ilmok, they see opposite spin structures
on $\tilde S$.  So $\psi_1$ and $\psi_2$ together have precisely one zero mode
on $S\times \tilde S$. Having an odd number of fermion zero modes means that
the partition function vanishes, and that this vanishing cannot be lifted
by insertions of local operators (a fermionic operator will not have an
expectation value once we average over spatial rotations).  It should
be interpreted as a kind of global anomaly.  We will argue in section 5.1
that  the $M5$-brane has such an inconsistency
on $S\times R$ whenever $R$ is not $\spinc$.

\bigskip\noindent{\it Analog For Type IIB}

Now let us briefly discuss the analogous issues in the other
case mentioned in the introduction.

Our goal is to compare topological effects in Type IIB and Type IIA
superstring theory on $S\times Y$, with $S$ a circle and
$Y$ a nine-dimensional
spin manifold.  But a shortcut along the
above lines is possible for the special case $Y=\tilde S\times Y'$, with 
$\tilde S$
another circle and $Y'$ an eight-dimensional spin manifold.
So we consider Type IIB superstring theory on $S\times \tilde S\times Y'$,
with the supersymmetric spin structure on the first factor.

\def\HP{{\bf HP}}
For illustration, we consider the case that $Y'={\bf HP}^2$.  The only
nontrivial cohomology group of this manifold  is 
$H^4({\bf HP}^2;\Z)=\Z$.  The generator is a self-dual four-form 
$\omega$ such that
\eqn\tomigo{\int_{{\bf HP}^1}\omega = 1,~~\int_{{\bf HP}^2}\omega\wedge\omega=1.}  Here ${\bf HP}^1$ is a linearly
embedded subspace of ${\bf HP}^2$ and generates $H_4({\bf HP}^2;\Z)$.
$\HP^2$ is a spin manifold, so its first Pontryagin class $p_1$ is
divisible by 2, and $\lambda=p_1/2$ obeys
\eqn\jomigo{\int_{\HP^1}\lambda = 1.}
In fact, $\lambda$ is just $\omega$.

We can repeat much of what we have already seen.
In compactification on $S\times \tilde S\times \HP^2$, with the last
factor much smaller than the first two, 
the chiral four-form $C_4$ of Type IIB superstring theory reduces at
long distances (via an ansatz $C_4=\omega\phi$) to a chiral
scalar $\phi$ on $S\times \tilde S$.  $\phi$ can be expressed in terms of
free fermions, and by the same reasoning
as above, if we regard $\tilde S$ as the ``time'' direction, then the
conserved charge
\eqn\hubo{q=\oint_{S}{d\phi\over 2\pi}}
takes half-integral values. 
One can think of $q$ more microscopically as
\eqn\trubo{q=\oint_{S\times \HP^1}{G_5\over 2\pi}}
where $G_5=dC_4$ is the gauge-invariant self-dual five-form of Type IIB.

Now we consider a $T$-duality transformation on the first circle $S$.  
This maps
Type IIB superstring theory to Type IIA, and the modes of $G_5$ that
appear in the integral in \trubo\ are mapped to $G_4$, the Ramond-Ramond
four-form field strength of Type IIA.  In the Type IIA description,
$q$ becomes
\eqn\orubo{q=\int_{\HP^1}{G_4\over 2\pi}.}
Thus, to account for the half-integrality of $q$ from the Type IIA
point of view,
we must explain why $G_4$ has half-integral periods in
this situation.

But this is a consequence of \jomigo.  The general formula is indeed
\ugwitten\ 
\eqn\olpo{\int_U{G_4\over 2\pi}={1\over 2}\int_U\lambda +{\rm integer},}
for any four-cycle $U$ in a Type IIA spacetime.  
In view of \jomigo, this is equivalent to half-integrality of $q$.
Thus, we have succeeded,
in this situation, in reconciling the gravitational shift in the
quantization law of the four-form in Type IIA  with the
subtleties of the self-dual five-form of Type IIB.  

For a more complete study of these problems, where we compactify on
only one circle and not two, we need to delve into the theory
of chiral $p$-form fields for $p>0$.  This will be the subject
of sections 4 and 5.  But first we must recall some additional
aspects of the quantum mechanics of self-dual $p$-forms, starting with
the one-form case.

\newsec{Quantum Mechanics Of Self-Dual $p$-Forms}

\nref\twitten{E. Witten, ``On $S$ Duality In Abelian Gauge Theory,''
hep-th/9505186.}%

Before looking at our specific problem, we need some more background
on chiral $p$-forms.

In constructing the quantum mechanics of an ordinary (not self-dual)
$p$-form field on a manifold $M$, one sums over all periods in
$H^p(M;\Z)$.  That is not so for a self-dual $p$-form.

In fact, it is impossible to impose any classical quantization law
at all on the periods of a self-dual $p$-form.  To illustrate this,
let $\Sigma$ be a two-torus constructed as $\C/\Lambda$, where $\C$ is
the complex $z$-plane, and
$\Lambda$ is a lattice generated by complex numbers 1 and $\tau$
(with ${\rm Im}\,\tau>0$).  Let $A$ be a cycle in $\Sigma$ that
lifts in $\C$ to a path from 0 to 1, and let $B$ be a cycle that lifts
to a path from 0 to $\tau$.
Let $\lambda$ be a self-dual one-form.
Then $\lambda=c\,dz$ for some complex constant $c$.  
If we want, for example, $\int_A\lambda/2\pi$ to be integral,
we need $c\in 2\pi \Z$, while requiring $\int_B\lambda/2\pi$ to
be integral puts an entirely different condition on $c$.

What happens instead is that a self-dual $p$-form must be treated
quantum mechanically; one cannot treat its periods classically.
The partition function of such a field is written as a sum over only
half the periods.
For illustration, let us consider an example \godol\ that
is extremely important in string theory: a collection of $8k$ chiral
bosons $\phi_i$ in two-dimensions, for some integer $k$, associated with
an even unimodular lattice $\Gamma$ with positive definite
intersection form $(~,~)$.  We set $\lambda_i=d\phi_i$.
The partition function in
genus one is as follows.  Let  $\Sigma$ be as above
and  $q=\exp(2\pi i\tau)$.  Then the partition function of
the chiral boson theory on $\Sigma$ is
\eqn\olop{Z(q)={\sum_{w\in \Gamma} q^{(w,w)/2}\over \eta(q)^{8k}}}
with $\eta$ the Dedekind eta-function.
In this formula, the partition function is constructed as a sum over a
single set of periods -- the periods $w_i=\int_A\lambda_i/2\pi$, which are
the components of a single lattice vector $w\in\Gamma$.  In a Hamiltonian
framework with $A$ regarded as the spatial cycle and $B$ as time,
the $A$-periods label the winding (or by self-duality
the momentum) states; the theta function in the numerator
of \olop\ comes from the sum over these states.  Of course,
the choice of the particular cycle $A$ is not uniquely determined.
  The partition function is 
$SL(2,\Z)$-invariant; by an $SL(2,\Z)$ transformation, one could
replace the cycle $A$ by $nA+mB$ for any relatively prime integers $n,m$.

Intuitively, we may think of two periods $\int_A\lambda$ and 
$\int_{A'}\lambda$ as commuting if and only if the intersection
number $A\cap A'$ is zero.  There is no way to simultaneously measure
noncommuting periods.  The partition function is constructed as a sum over
a maximal set of commuting periods.  

The example relevant to the present paper is slightly more subtle:
it is the case that  the chiral bosons $\phi_i$ are derived from
a lattice $\Gamma$ that is unimodular, but not even.  In fact, the
prototype for us is a single chiral boson at the free fermion radius,
that is to say $\Gamma$ is a one-dimensional lattice generated by a vector
$\omega$ with $(\omega,\omega)=1$.  In this case, there is not a single
partition function; rather (as is apparent from the description by free
fermions) there is a partition function for each choice of spin
structure.  It is instructive to examine these partition functions.
They are conveniently written in terms of standard functions as \eqn\kloop{
Z\left[\matrix{\theta \cr\phi\cr}\right](z|\tau)={\vartheta
\left[\matrix{\theta \cr\phi\cr}\right](z|\tau)\over \eta(\tau)},}
where $\theta$ and $\phi$ are 0 and 1/2 and $z$ is an extra variable
included to represent the coupling to a background gauge field.
The partition function in the absence of this field is obtained by
setting $z=0$.  The functions in the numerator on the right hand side
are called theta functions with characteristics.
They are explicitly
\eqn\oloop{\eqalign{\vartheta \left[\matrix{0 \cr 0\cr}\right](z|\tau)
   &  = \sum_{n\in \Z}q^{n^2/2}\exp(2\pi in z)      \cr                                    \vartheta\left[\matrix{0 \cr 1/2\cr}\right](z|\tau)& =\sum_{n\in\Z}
 (-1)^nq^{n^2/2}\exp(2\pi i n z)           \cr
 \vartheta\left[\matrix{1/2 \cr 0\cr}\right](z|\tau)& =\sum_{n\in \Z+1/2}
q^{n^2/2} \exp(2\pi i nz)            \cr
 \vartheta\left[\matrix{1/2 \cr 1/2\cr}\right](z|\tau)  & =i\sum_{n\in \Z+1/2}
(-1)^{n+1/2}q^{n^2/2}\exp(2\pi i n z)  .  \cr}}
We have written these theta functions as sums over the $A$-period
$n=\int_A d\phi/2\pi$.  By $SL(2,\Z)$, one could instead write
each of these theta functions as a sum over any other chosen period
of $d\phi$.  While $\vartheta\left[\matrix{1/2 \cr 1/2\cr}\right]$,
which corresponds to the odd spin structure, is $SL(2,\Z)$-invariant
(up a a $c$-number multiple that reflects the modular weight plus
an anomalous phase),
the others are permuted by $SL(2,\Z)$, so if one chooses to write
$\vartheta\left[\matrix{0 \cr 0\cr}\right]$, for example, with a different
choice of the period, one might have to use the formula for
$ \vartheta\left[\matrix{1/2 \cr 0\cr}\right]$.

In constructing the theta function as a sum over the values of the
$A$-period $n$,
this period is  integral for $\theta=0$ and
half-integral for $\theta=1/2$.  
Therefore the answer to the question of whether a given period of
the self-dual one-form is integral or half-integral depends on the choice
of theta function.  On the other hand, $\phi$ determines the sign factors
in the sum over the $A$-periods.  A configuration with a given
value of the $A$-period 
$n$ is weighted by a sign $+1$ if $\phi=0$ and by a sign
$(-1)^n$ (or $(-1)^{n+1/2}$ if $n$ is half-integral) if $\phi=1/2$.

\def\Om{\Omega}
Now, we want to describe the theta functions in a way that generalizes
to higher genus surfaces and also to self-dual $p$-forms of $p>1$.
We will define a $\Z_2$-valued function $\Om(x)$ on the lattice  $\Lambda$
as follows.\foot{In \witten, this function was called $H(x)$, but I want
to avoid notational
clashes with the three-form field $H$ of string theory and $H^i(M)$
for cohomology groups.}  For the lattice points 1 and $\tau$, we set
\eqn\olbno{\Om(1)=(-1)^{2\phi}, ~~\Om(\tau)=(-1)^{2\theta}.}
We extend $\Om$ to a function on the whole lattice by requiring
\eqn\hulbun{\Om(x+y)=\Om(x)\Om(y)(-1)^{(x,y)},}
where $(x,y)=-(y,x)$ is the intersection form on the lattice $\Lambda$.
For example, this definition gives 
\eqn\olo{\Om(1+\tau)=-\Om(1)\Om(\tau),}
since $1$ and $\tau$ correspond to the cycles $A$ and $B$,
whose intersection number is 1.
\hulbun\ is the basic formula.  Theta functions are in natural one-to-one
correspondence with $\Z_2$-valued functions on the lattice that obey
this relation.  Given $\Om$, the characteristics $\theta,\phi$ are extracted
from \olbno\ and used to write the explicit formulas for the theta
functions that we gave above.

Let $\Lambda_1$ and $\Lambda_2$ be, respectively, the sublattices of
$\Lambda$ generated by 1  and by $\tau$; we call these the $A$-lattice
and the $B$-lattice.
As we saw above, a configuration with $A$-period $n$ contributes
to the theta function (in the representation of that function
 as a sum over the $A$-periods) with
a sign $1$ or $(-1)^n$ depending on $\phi$.  \olbno\ means that $\Om(x)$
for $x$ in the $A$-lattice is simply the sign factor with
which a configuration of $A$-period  $n=x$ (or $n=x+1/2$)
 contributes to the theta function.
Likewise, we saw above that $\theta$ determines whether the $A$-periods
are integral or half-integral, and thus this is determined by $\Om(x)$
for $x$ in the $B$-lattice.

The classification of theta functions by $\Z_2$-valued functions $\Om(x)$
extends beyond the genus one case that we have just considered:
level one theta functions of any lattice $\Lambda$ with unimodular
antisymmetric form $(~,~)$ and a metric for which this form is positive
and of type $(1,1)$ are classified by functions $\Om$ obeying \hulbun.
This fact has
a differential-geometric explanation that was reviewed in \witten.
(The basic idea is that 
such an $\Om$ determines a line bundle over $\Sigma$; this line bundle
has up to constant multiples a unique holomorphic section which is the
theta function.)  For our present purposes, we will
simply note that the functions $\Om$ that obey \hulbun\ transform
under $SL(2,\Z)$ the same way that theta functions do.  In
this assertion, the sign factor $(-1)^{(x,y)}$ in \hulbun\ is essential.
For example, the theta function 
$ \vartheta\left[\matrix{1/2 \cr 1/2\cr}\right]$
associated with the odd spin structure is $SL(2,\Z)$-invariant,
so it must be associated with a function $\Om(x)$ that is likewise
$SL(2,\Z)$-invariant.  Since $\theta=\phi=1/2$, this theta function
has $\Om(1)=\Om(\tau)=-1$.  As $SL(2,\Z)$ can map the lattice points
 1 or $\tau$ to $1+\tau$,
it follows that $\Om(1+\tau)$ must equal $-1$, which is what we get from \olo.

To write the four theta functions by  explicit formulas as in \oloop\
requires a choice of $A$-lattice.  Some more information is needed,
though, because the choice of $A$-lattice is invariant under $\tau\to\tau+1$,
but this operation permutes the theta functions in a non-trivial
fashion.  If one is also given a choice of $B$-lattice (and thus
essentially the
basis $(1,\tau)$ for the lattice $\Lambda$), this is more than
enough information to enable the writing of the explicit formulas in 
\oloop.  (For that, it is enough to know the $B$-cycles mod 2.)
If one has chosen both the $A$-lattice and the $B$-lattice, then one
has an explicit $SL(2,\Z)$ transformation $\tau\to -1/\tau$ that
exchanges them.  It exchanges $\theta$ and $\phi$, and thus exchanges
a half-integral shift in the value of the $A$-period $n$ with a sign
factor by which the different values of the $A$-period are weighted.

\bigskip\noindent{\it Generalization}

Now let us consider the generalization to a self-dual $p$-form field $G_p$,
of $p$ possibly bigger than 1, on a $2p$-dimensional
manifold $M$.  (For a detailed treatment via holomorphic factorization
of the partition function of a non-chiral theory, see \hns.)
The periods
take values in $\Lambda=H^p(M;\Z)$, which for simplicity we will assume
to be torsion-free.  Thus $\Lambda$ is a lattice, with an antisymmetric
bilinear form $(~,~)$ of determinant 1 that is given by the intersection
pairing on $M$.  If $\Lambda$ has rank $2g$, then it has
 has $2^{2g}$ distinguished theta functions
$\vartheta
\left[\matrix{\theta \cr\phi\cr}\right](z|\tau)$ that we will
introduce momentarily.  The partition
function of $G_p$ is $\vartheta
\left[\matrix{\theta \cr\phi\cr}\right](z|\tau)/\Delta$, where $\Delta$
(analogous to $\eta(\tau)$ in \kloop) is uniquely determined from
the non-zero modes of $G$.  The subtlety comes from the choice of
theta function in the numerator.

As in the case of a one-form field, the periods of $G$ are not
all simultaneously measureable.  The best that one can do is
to pick a maximal sublattice $\Lambda_1$ consisting of mutually
``commuting'' periods.  $\Lambda_1$ is a lattice of $A$-periods,
that is, it is a half-dimensional sublattice of $\Lambda$ such
that $(x,y)=0$ for $x,y\in\Lambda_1$.  It is convenient, though
not necessary, to pick also a complementary lattice $\Lambda_2$
of $B$-periods.  Thus, $\Lambda=\Lambda_1\oplus \Lambda_2$, and 
$(x,y)=0$ for $x,y\in\Lambda_2$.  Picking the $B$-periods and $A$-periods
gives an explicit period matrix $\tau_{ij}=\tau_{ji}$, $i,j=1,\dots,g$
for the lattice $\Lambda$.

Once the $A$-cycles and $B$-cycles are fixed, one can write an
explicit formula for the theta functions.  One picks a half-lattice
vector $\theta\in \half\Lambda_1/\Lambda_1$, and a half-lattice vector
$\phi\in \half\Lambda_2/\Lambda_2$.
The theta function with characteristics $\theta$, $\phi$ is then
\eqn\julki{\vartheta
\left[\matrix{\theta \cr\phi\cr}\right](z_i|\tau)=\sum_{n\in \Lambda_1+\theta}
\exp\left(i\pi\sum_{ij}n^in^j\tau_{ij}+2\pi i n^i(z_i+\phi_i)\right).}
The $z_i$ are parameters that measure the coupling to a background
$p$-form potential; the partition function is obtained by setting $z_i=0$
(and dividing by $\Delta$).  

From \julki, we see that if we write
the theta function as a sum over $A$-periods, then the $A$-periods
are shifted from integers by $\theta\in \half\Lambda_1/\Lambda_1$.
But the sign factor in the sum over $A$-periods is determined by $\phi$.

As in the $g=1$ case that we discussed first, the theta functions
are most naturally classified by a $\Z_2$-valued function $\Om(x)$ on
the lattice $\Lambda$
that obeys the fundamental relation
\eqn\uccub{\Om(x+y)=\Om(x)\Om(y)(-1)^{(x,y)}~{\rm for~all}~x,y\in\Lambda.}
Given such a function, one defines the characteristics $\theta, \phi$ by
\eqn\defto{\eqalign{\Om(x)&=(-1)^{2(x,\phi)}~{\rm if}~x\in \Lambda_1\cr
                    \Om(x)&=(-1)^{2(x,\theta)}~{\rm if}~x\in\Lambda_2,\cr}}
and then the theta function can be defined by the formula in \julki.
As mentioned above, there is also a more intrinsic procedure to go from
$\Om$ to the theta function (use $\Om$ to construct a line bundle and
take its holomorphic section).

Combining the above definitions,
we can see how the theta function depends on $\Om$.
For $x\in \Lambda_1$, $\Om(x)$ is a sign factor in the sum over
$A$-periods, and for $x\in \Lambda_2$, $\Om(x)$ controls the non-integrality
of the $A$-periods. This generalizes what we explained above for $g=1$.

\bigskip\noindent{\it Specializing to $M=S\times F$}

In general, this formalism is somewhat abstract, partly because for
a general $2p$-dimensional manifold $M$, there is no particularly
nice choice of $A$-periods and $B$-periods.  Nice choices
do exist if $M=S\times F$, with $S$ a circle and $F$ a manifold
of dimension $2p-1$.  This case is our focus in the present
paper.  The theory of a self-dual $p$-form $G_p$ on such an $M$ reduces
at low energies on $F$ to a theory of an ordinary $p$-form $G_p'$ with
no self-duality, or
(after a duality transformation) to a theory of an ordinary
$(p-1)$-form  $G_{p-1}'$.  Correspondingly, the cohomology of $M$ splits
as \eqn\jugoc{\eqalign{H^p(M;\Z)=&H^p(F;\Z)\oplus H^1(S;\Z)\otimes_\Z H^{p-1}(F;\Z)\cr
=&H^p(F;\Z)\oplus H^{p-1}(F;\Z).\cr}}
We take $\Lambda=H^p(M;\Z)$, so that a partition function of
$G_p$ on $M$ is a theta function of $\Lambda$, and we set $\Lambda_1=H^p(F;\Z)$,
$\Lambda_2=H^{p-1}(F;\Z)$.

We take the lattice
of $A$-periods to be $\Lambda_1$ and the lattice of $B$-periods to
be $\Lambda_2$.  A theta function for $\Lambda$ can be constructed
either as a sum over $A$-periods -- corresponding to the representation of the
theory on $F$ in terms of $G_p'$ -- or as a sum over $B$-periods --
corresponding to the representation of the theory on $F$
in terms of $G_{p-1}'$.  

No matter which representation one uses, the theta function of $\Lambda$
is determined by a choice of a suitable function $\Om(x)$ on $\Lambda$.
How to construct such a function for the self-dual $p$-form fields mentioned
in the introduction was explained in \witten.  Once $\Om$ is selected,
its restriction to $\Lambda_1$ determines a sign factor in the sum over
periods if one uses the description of the theory in terms of $G_p'$,
or a nonintegrality of the $G_{p-1}'$ periods in the other description.
Conversely, the restriction of $\Om$ to $\Lambda_2$ determines the
nonintegrality of $G_p'$ periods, or a sign factor in the sum over $G_{p-1}'$.

The fact that a sign factor on one side becomes, after duality,
a nonintegrality of the periods in the other description can also
be explained more microscopically in terms of a Feynman path integral
representation of the $G_p'$ and $G_{p-1}'$ theories.  More generally,
in $d$ dimensions, a phase factor in a theory with $k$-form curvature
$G_k$ is always translated after duality to a shift in periods of a dual
field $G_{d-k}$.  This holds for all $d$ and $k$.
A path integral derivation of this fact can be found
(for $d=4$, $k=2$, but the general case is not essentially different)
in section 4.2 of \twitten.

\newsec{Systematic Analysis Of Type II Case}

In attempting a systematic treatment, using the framework
of \witten, of the problems mentioned in the introduction,
we will begin with the second problem -- understanding the shifted
quantization law of the Type IIA four-form from the quantum mechanics
of the self-dual five-form of Type IIB.  This case involves
fewer technicalities.

\subsec{Outline}

In Type IIB theory on a ten-dimensional spin manifold $X$, we have a four-form potential $C_4$ with a self-dual
curvature five-form $G_5$.  If we could omit the self-duality condition,
and we impose conventional Dirac quantization, then the $C_4$-fields
are classified topologically by a class $x\in H^5(X;\Z)$.
Here $x$ is represented in de Rham cohomology by $G_5/2\pi$.  We sometimes
write informally $x=[G_5/2\pi]$.

For an ordinary four-form field, we would construct the partition function
by summing over all choices of $x$ (and for each choice of $x$, integrating
over all possibilities for $C_4$).  For a four-form
of self-dual curvature, we do not sum independently over all values of $x$.
Rather, as discussed in section 3 above
and in \witten, we construct the partition function
in terms of a theta function on $\T = H^5(X;U(1))$, which,
if there is no torsion in the cohomology of $X$, is the torus $H^5(X;\R)/H^5(X;\Z)$.  The theta function, as we discussed in 
section 3, 
is constructed by summing over a maximal set of ``commuting'' periods.
\foot{In general, $H^5(X;U(1))$ has components labeled by the torsion subgroup of $H^6(X;\Z)$; each component is a torus.
This refinement  will not be essential in our present discussion, and I suspect that the torsion can be fully taken into account only if one works
with $K$-theory rather than cohomology, a task that we initiate in section
4.3 below.  Note that in what follows, we write the product of
differential forms as a wedge product, denoted $\wedge$, and the product
of integral cohomology classes as a cup product, denoted $\cup$.}

To construct the theta function, as explained in \witten\ and
in section 3, we need a function
$\Om(x)$ from $H^5(X;\Z)$ to the group $\Z_2=\{\pm 1\}$, obeying
\eqn\tufub{\Om(x+y)=\Om(x)\Om(y)(-1)^{x\cdot y}}
for all $x,y\in H^5(X;\Z)$.  Here $x\cdot y$ is the intersection
pairing $\int_X x \cup y$.  The function $\Om$ is needed to determine
a line bundle on $\T$, a suitable section of which is the theta function.
It is convenient to write $\Om(x)=(-1)^{h(x)}$, where $h(x)$ is
an integer-valued function that is defined modulo two.

In what follows, we will study the function $h(x)$ for the case that
$X=S\times Y$, with $S$ a circle endowed with a spin structure of
unbroken supersymmetry (that is,  a non-bounding spin structure) and $Y$ an arbitrary nine-dimensional spin-manifold.
We will find that if $x$ is an element of $H^5(Y;\Z)$,
then 
\eqn\milok{h(x)=\int_Y \lambda \cup x,}
where $\lambda$ is the integral characteristic class such that $2\lambda
=p_1(Y)$.  $H^5(X;\Z)$ is generated by $H^5(Y;\Z)$ together
with elements of the form
 $a\cup w$ with $a$ a generator of $H^1(S;\Z)=\Z$
and $w\in H^4(Y;\Z)$.  So the function $\Om(x)$ would be completely
determined by \milok, \tufub, and a knowledge of $h(a\cup w)$.
It does not seem that there is a formula for $h(a\cup w)$
as elementary as \milok.\foot{For example, \milok\ shows that $h(x)$
is independent of the spin structure for $x\in H^5(Y;\Z)$; but
examples such as $Y=\tilde S\times Y'$ (with $\tilde S$ another circle and $Y'$
an eight-manifold) show that $h(a\cup w)$ does depend on the spin structure.
In this case, for $w\in H^4(Y';\Z)$, $h(a\cup w)$ is equal to $\int_{Y'}
\lambda \cup w$ or 0 depending on whether one takes the supersymmetric
or nonsupersymmetric spin structure on $\tilde S$.   This can be seen
by the methods of section 2.}
It turns out, though, that \milok\ suffices to answer the question
raised in the introduction.

This comes about as follows.  The use of the
function $\Om(x)$ to determine the partition function on a general $X$ is
perhaps slightly esoteric.  But this function has a more down-to-earth
interpretation if $X$ is of the form $S\times Y$.  In this case, the
theory of the self-dual five-form $G_5$ on $X$ reduces on $Y$, at low
energies, to a theory of a five-form on $Y$ that obeys no self-duality
condition.  We will
call this field $G_5'$.  As $Y$ is nine-dimensional, the theory of
$G_5'$ is dual to a theory of a four-form field $G_4'$ on $Y$.
$G_4'$ and $G_5'$ are the curvatures of three-form and four-form potentials
$C_3'$ and $C_4'$.  The same theory on $Y$ can be described with
either $C_3'$ or $C_4'$ as the dynamical variable.

$\Om(x)$ has, as we have seen in section 3,
the following straightforward interpretation when
$X$ is of the form $S\times Y$: it is a factor that
must be included in the low energy path integral for the $G_5'$ field
on $Y$. To be more precise, if we regard the theory on $Y$ as the theory
of a five-form $G_5'$ with flux or characteristic class $x=[G_5'/2\pi]$, 
then performing the path integral involves summing over $x$.
The sum is weighted with a number of standard factors, such as an obvious
factor coming from the kinetic energy of the $C'_4$ field.
In addition, we must include in the sum over $x$ the sign factor
$\Om(x)$, which according to \milok\ is
\eqn\onox{\exp\left(i\pi\int_Y\lambda \cup x\right)=\exp\left(
\half i\int \lambda \wedge G_5\right).}

On the other hand, if we represent the theory on $Y$ by a four-form
$G_4'$ with flux or characteristic class $u=[G_4'/2\pi]$, then
performing the path integral involves summing over $u$.  The sign factor we
must include in the path integral is in this case
$\Om(a\cup u)$ (since the relation between $G_5$ and $G_4'$ is that the
appropriate part of $G_5$ is $a\wedge G_4$), 
and we will not determine $\Om(a\cup u)$.  In addition to this in general
unknown
sign factor, the path integral over $G_4'$ has another interesting
effect, which arises by duality from \onox.\foot{And, conversely to what
we are about to say, the undetermined sign factor $\Om(a\cup u)$ in the $G_4'$
theory will
by duality, if it is not trivial, induce a shift in the periods of $G_5'$.}
As we have discussed in section 3, and as was explained from a path
integral point of view in \twitten,
 a phase factor on
one side is converted by duality into a shift on the other side.
In the present instance, since the phase in  \onox\ is 
 $\half\lambda$ (times $G_5'$), 
it is converted by duality into a shift in the
periods of $G_4'$ by that amount: for any four-cycle $U\in Y$,
\eqn\ilboxo{\int_U{G_4'\over 2\pi} ={1\over 2}\int_U\lambda ~{\rm mod}~\Z.}

The last formula is essentially the result that we need.  The problem
posed in the introduction was to understand, starting with the quantum
mechanics of the Type IIB self-dual five-form $G_5$, 
the fact that for any four-cycle $U$ in a Type IIA spacetime $X$,
\eqn\pilboxo{\int_U{G_4\over 2\pi} ={1\over 2}\int_U\lambda ~{\rm mod}~\Z,}
where here $G_4$ is the Type IIA four-form. 
If $X=S\times Y$ with $S$ a circle,
then components of $\lambda$ with an index tangent to $S$ vanish
topologically (since $\lambda$ is a pullback from $Y$),
and the interesting case of \pilboxo\ is the case with $U$ a four-cycle
in $Y$.  Hence the interesting part of $G_4$ is the part with all indices
tangent to $Y$.
In the $T$-duality between Type IIB and Type IIA on $S\times Y$, this
part of $G_4$  is related to the part of
$G_5$ of the form $a\cup G_4$, with $a$ a generator of $H^1(S;\Z)$.  
So the relevant part of $G_4$ is the same as $G_4'$,
and \ilboxo\ is equivalent to the desired
relation \pilboxo.

In the next 
subsection, we will justify the crucial formula \milok. Then in  section 4.3,
we will propose a new description of $\Om(x)$ in $K$-theory which may
be more useful for understanding dualities and the role of torsion.

\subsec{  Evaluation Of $\Om(x)$}

First we recall from \witten\ the definition of $\Om(x)$ for a general
$X$.  We work on $Z=S'\times X$ where $S'$ is a circle with a Neveu-Schwarz
spin structure (that is, $S'$ is a spin boundary).  We fix a generator
$a'$ of $H^1(S';\Z)=\Z$.  For $x\in H^5(X;\Z)$, we set $z= a' \cup x\in
H^6(S'\times X;\Z)$.  Now, if $W$ is any twelve-dimensional spin manifold
with boundary $Z$ over which $z$ extends,\foot{How to generalize
the discussion if $W$ does not exist is discussed in \witten.
We also give more general definitions of $\Om(x)$
below and in section 4.3.}
 we set
\eqn\loko{h(x)=\int_W z\cup z}
and $\Om(x)=(-1)^{h(x)}$.
For $\Om(x)$ to be well-defined,
$h(x)$ must be independent modulo 2 of the choice of  $W$.  This is so because
for a {\it closed} twelve-dimensional spin manifold $W$ (that is,
one whose boundary vanishes),
$\int_W z \cup z$ is even for any $z\in H^6(W;\Z)$. (A proof
of this assertion is cited in a footnote in section 4 of \witten.)

We could calculate more conveniently if we did not have to require
$W$ to be a spin manifold.  However, if we try to use the definition
\loko\ without requiring $W$ to be spin, $h(x)$ would not be well-defined
modulo 2 because $\int_W z\cup z$ is not necessarily even on a general
twelve-manifold $W$.  There is, however,
an analogous quantity that is even in general; it 
is $\int_W(z\cup z+v\cup z)$, where $v$ is the (six-dimensional) Wu class 
of $W$.  $v$ can be expressed in terms of Stieffel-Whitney classes; for
our purposes, we can assume that $W$ is orientable, in which case
the relation is $v=w_2\cup w_4$.  Thus, we are tempted to generalize
the definition of $h(x)$ to
\eqn\polko{h(x)= \int_W(z\cup z+w_2\cup w_4\cup z)}
where $W$ is now required only to be oriented.  (To make sense of the
second integral, $z$ is reduced mod 2 and the integral is understood
in terms of the cup product and integration in mod 2 cohomology.)

Some care is needed here.  Though the right hand side of \polko\ is
indeed even if the boundary of $W$ vanishes, some subtlety enters in
defining the integral when $W$ has a nonzero boundary.  
An integral such as \polko\ is not a topological invariant
 on a manifold with
boundary unless the class that is being integrated is trivialized on the
boundary; and even if it is, the integral depends on the choice
of a trivialization on the boundary.  (At the level of differential
forms, this statement means that an integral $\int_W\Theta$,
where $\Theta$ is a twelve-form, is not  necessarily
invariant under $\Theta\to \Theta
+d\Lambda$ if $\Lambda$ is nonvanishing on the boundary.)
In the case of \polko, if we understand $z$ near the boundary $Z$
of $W$ to be a pullback from $Z$, then $z \cup z$ vanishes
near the boundary for dimensional reasons, and this trivialization is natural.
We need more care with the term $w_2 \cup w_4\cup z$.

As $z$ and $w_4$ are both in general nonzero near the boundary, the only
reason that $w_2\cup w_4\cup z$ vanishes near the boundary is that $w_2$
does, that is, the boundary manifold $Z$ is spin.  A trivialization
of $w_2$ near the boundary is a choice of spin structure on $Z$,
and hence we will have to use the spin structure of $Z$ in defining the
integral $\int_W w_2\cup w_4\cup z$ even though at first sight the
integral appears not to depend on a choice of spin structure.
A rather down-to-earth way to build in the spin structure of $Z$ is
to restrict to the case that $W$ is a $\spinc$ manifold, with a $\spinc$
structure that extends the spin structure on $Z$.  The $\spinc$ structure
on $W$ determines an integral lift\foot{A $\spinc$ structure determines
a $\spinc$ bundle that is informally ${\cal S}\otimes {\cal L}^{1/2}$
where ${\cal S}$ is the spin bundle  and ${\cal L}$ is a line bundle with $c_1({\cal L})$ congruent
to $w_2(W)$ mod 2.  If $W$ is not spin, neither ${\cal S}$ nor
${{\cal L}^{1/2}}$ exists separately, but ${\cal S}\otimes {\cal L}^{1/2}$
and ${\cal L}$ do.
$\alpha$ is defined as $c_1({\cal L})$.}
 $\alpha$ of $w_2(W)$ that is supported
away from the boundary of $W$; to be more
precise, it determines an element $\alpha$ of the relative cohomology group
$H^2(W,\partial W;\Z)$ that reduces to $w_2(W)$ mod 2.  Moreover,
if $W$ and $W'$ are two $\spinc$ manifolds with (oppositely oriented)
boundary $Z$ and $\spinc$ structures that extend the spin structure of $Z$,
then upon gluing together $W$ and $W'$ to make a closed twelve-dimensional
$\spinc$ manifold $\bar W$, the corresponding classes $\alpha$ and $\alpha'$
glue together to an integral lift $\bar\alpha$ of $w_2(\bar W)$. 
In fact, $\bar\alpha$ is derived from the $\spinc$ structure on $\bar W$
that is obtained by gluing those on $W$ and $W'$.   (Gluing $\alpha$ and
$\alpha'$ to make an integral lift $\bar\alpha$ of $w_2(\bar W)$
would not work for arbitrary integral lifts $\alpha$
and $\alpha'$ of $w_2(W)$ and $w_2(W')$; that is why it is important
to derive $\alpha$ and $\alpha'$ from $\spinc$ structures that extend
that of $Z$.)

Finally, we get our more general definition of $h(x)$:
\eqn\tufalo{h(x)=\int_W\left(z\cup z+\alpha\cup w_4\cup z\right).}
This is well-defined mod 2 because it is even for a closed twelve-dimensional
$\spinc$ manifold $\bar W$.

\bigskip\noindent{\it Evaluation For $X=S\times Y$}

We are ready to compute for the case that $X=S\times Y$,
with the supersymmetric (or  non-bounding) spin structure on $S$.
We have $Z=S'\times X=S'\times S\times Y$.

We want to evaluate $h(x)$ where $x$ is an element of $H^5(Y;\Z)$.
We have $z=a'\cup x$.  To compute $h(x)$, we should write $Z$ as a boundary
of a $\spinc$ manifold $W$ over which $z$ extends.  We could try
to take $W=D'\times S\times Y$, where $D'$ is a two-dimensional disc
with boundary $S'$.  This is not convenient because $a'$ does not
extend over $D'$.  Instead, we let $D$ be a disc with boundary $S$,
and set $W=S'\times D\times Y$.
The spin structure of $S$ does not extend over $D$ as a spin structure,
but it extends as a $\spinc$ structure with 
\eqn\uffgo{\int_D\alpha=1.}
As $z$ is a pullback from $S'\times Y$, it extends over $W=S'\times D\times Y$
as such a pullback.
Now we can evaluate \tufalo.  On dimensional grounds, since  $z$ is
pulled back from $S'\times Y$, $z\cup z=0$.  So we need only consider
the integral over $S'\times D\times Y$
of $\alpha\cup w_4\cup z = \alpha \cup w_4 \cup a'\cup
x$.  The integral is easily done because all factors are pullbacks from
one of the factors in $S'\times D\times Y$ ($a'$ from $S'$, $\alpha$ from
$D$, and the others from $Y$).  Using \uffgo\ and $\int_{S'}a'=1$, we get
\eqn\nuffgo{h(x)=\int_Yw_4\cup x=\int_Y\lambda\cup x,}
where the two expressions are equivalent because on the spin manifold $Y$,
$\lambda$ is an integral lift of $w_4$.  This is the promised
formula \milok.

\subsec{ $K$-Theory Definition Of $\Om(x)$}

We have performed this computation in a framework \witten\ in which
$\Om(x)$ is defined as a function on middle-dimensional cohomology
of Type IIB.  For two reasons, it seems that the definition should
be reformulated in $K$-theory:

(1) In view of $T$-dualities which relate the Ramond-Ramond (RR) forms
of different dimensions, and relate Type IIB to Type IIA, it seems
unnatural to have a special formalism which only applies to the middle-dimensional RR form for Type IIB, 
and does not apply at all for Type IIA.
If we define $\Om(x)$ in $K$-theory, this will automatically include
all of the RR forms of all even or all odd dimension, and may give
a $T$-dual formalism.

(2) In view of what we now know about the RR fields, it seems
unlikely that one can correctly take into account the torsion
part of the RR fluxes without using $K$-theory instead
of cohomology.

The rest of this section is devoted to an attempt to give a $K$-theory
definition of $\Om(x)$.

For Type IIA at the level of differential forms, the total RR field $G=G_0+G_2+G_4+\dots$ is a sum of differential forms of all even orders.  
For Type IIB, one has instead a sum $G=G_1+G_3+ \dots$  
of differential forms of all odd orders.  In passing to $K$-theory,
we will assume that for Type IIA, the RR flux should be regarded
as an element $x\in K(X)$.  For Type IIB, it should be regarded as
an element $x\in K^1(X)$.

\def\ch{{\rm ch}}
We will first define a $\Z_2$-valued function $\Om(x)=(-1)^{h(x)}$ for
$x\in K(x)$, that is, for Type IIA.  We want
\eqn\ucucb{\Om(x+y)=\Om(x)\Om(y)(-1)^{(x,y)},}
where $(x,y)$ should be an integer-valued bilinear form on $K(X)$
that generalizes the intersection pairing on cohomology.  Moreover,
we want $(x,y)=-(y,x)$, so that $\Om(x)$ can be used to define a line
bundle on a torus $K(X;\R/\Z)/K(X;\Z)$ (by analogy with what is done
for the middle-dimensional cohomology in \witten).
A suitable definition is given by index theory.  For any $w\in K(x)$, let 
\eqn\loppo{i(w)=\int_X\hat A(X)\ch(w)}
be the index of the Dirac operator with values in $w$.  In ten-dimensions,
the only terms in $\ch(w)$ that contribute are terms of degree $4k+2$ for
some integer $k$. These terms are odd under $w\to \bar w$
(complex conjugation of the bundle) so 
\eqn\guflod{i(w)=-i(\bar w).}  Then we set
\eqn\gloppo{(x,y)=i(x\otimes\bar y),}
which obeys $(x,y)=-(y,x)$ by virtue of \guflod.  This pairing
vanishes if $x$ or $y$ is torsion; it can be proved
that on $K(X)$ mod torsion, it is unimodular.

There is one more thing we should know about index theory in ten dimensions.
If $w$ is a real bundle, then $i(w)=0$ because of \guflod.  But there
is nonetheless a natural invariant of $w$ that can be defined using
index theory.  This  is the ``mod 2 index,'' the number of positive
chirality zero modes of the Dirac operator with values in $w$, modulo
two \ref\atsin{M. F. Atiyah and I. M. Singer, ``The Index Of Elliptic
Operators: V,'' Ann. Math. {\bf 93} (1971) 139.}.  
We will call this $j(w)$.  There is in general no elementary
formula for $j(w)$.  But if the complexification of $w$ is
of the form $x\oplus \bar x$ for some complex bundle $x$, then
\eqn\ilipo{j(w)=i(x)~{\rm mod}~2.}
In fact, $i(x)=n_+(x)-n_-(x)$, where $n_+(x)$ and $n_-(x)$ are respectively
the number of positive and negative chirality zero modes with values in
$x$.  Since in ten dimensions, complex conjugation reverses the
chirality, we have $n_-(x)=n_+(\bar x)$, so modulo
2 we have $i(x)=n_+(x)+n_+(\bar x)=n_+(w)=j(w)$.

We now can define $h(x)$, and hence $\Om(x)$, for Type IIA.
We simply set
\eqn\ifalot{h(x)=j(x\otimes \bar x).}
We must verify \ucucb.  If $z=x\oplus y$, then 
$z\otimes \bar z=x\otimes \bar x \oplus y\otimes \bar y
\oplus w$, with $w=x\otimes \bar y \oplus y\otimes \bar x$.
So \eqn\grifo{\eqalign{
h(x+y)=&j(z\otimes \bar z)=j(x\otimes \bar x)+j(y\otimes \bar y)
+j(w)\cr =&h(x)+h(y)+i(x\otimes \bar y)=h(x)+h(y)+(x,y),\cr}} as required.

We also want the analogous definition for Type IIB.  In this
case, we want to define a suitable function $\Om(x)$ for $x\in K^1(X)$.
We interpret $K^1(X)$ as $\tilde K(X\times \S^1)$, the subset of
$K(X\times\S^1)$ consisting of elements that are trivial if restricted
to $X$.  For $x,y\in K^1(X)$, we have $x\otimes \bar y\in K^2(X)=
\tilde K(X\times\S^1\times \S^1)$, and we define
\eqn\oglok{(x,y)=\int_{X\times\S^1\times \S^1}\hat A(X\times\S^1\times \S^1)\ch(x\otimes\bar y).}
This integer-valued function again obeys $(x,y)=-(y,x).$

Now we want to define $\Om(x)$.  Here there is a slight subtlety.
The element $x\otimes \bar x$ of $\tilde K(X\times \S^1\times \S^1)$
is replaced by its complex conjugate if one exchanges the two $\S^1$'s.
In addition, it is trivial if restricted to $X\times \S^1\times p$
or $X\times p\times \S^1$, with $p$ a point in one of the $\S^1$'s.  These
properties ensure that $x\otimes\bar x$ can be interpreted as
an element of $KR(X\times \S^2)$, where the real involution used
in defining $KR$ is a reflection of one coordinate of $\S^2$.
(By collapsing $\S^1\times p$ and $p\times \S^1$, one maps $\S^1\times \S^1$
to $\S^2$; the map that exchanges the two factors of $\S^1\times \S^1$
becomes a reflection of one coordinate in $\S^2$.)  By the
periodicity theorem of $KR$ theory \ref\atikr{M. F. Atiyah, ``$K$-Theory
And Reality,'' Quart. J. Math., Oxford (2) {\bf 17} (1966) 367.}, $KR(X\times \S^2)$
is the same as $KO(X)$.  So $x\otimes \bar x$ maps to an element
$w\in KO(X)$, and we define $h(x)=j(w)$.  The proof of \ucucb\
is rather as before.                                                                                                                                                                                                                                                                                                                                                                                                                                                                                                                                                                                                                                                                                                                                                                                                                                                                                                                                                                                                                                                                                                                                                                                                                                                                                                                                                                                                                                                                                                                                                                                         
\newsec{Systematic Analysis For $M5$-Brane}

In this section, we will carry out an analysis of the other
problem mentioned in the introduction --  the relation of the $M5$-brane
to the $D4$-brane -- analogous to what we have seen in section 4
for Type IIA/IIB.  The discussion will proceed in the following
stages: first we will summarize results; then we will compute
by hand; then we will place the computation more systematically
in the framework of \witten.

\subsec{Outline}

Let $V$ be the worldvolume of an $M5$-brane in an $M$-theory
spacetime $M$.  In general, $V$ is oriented, but perhaps not spin.

The subtle part of the quantum mechanics of the $M5$-brane is
to quantize the chiral two-form, which has a characteristic class
$x\in H^3(V;\Z)$.  The general framework for doing so is analogous
to what we summarized in the last section.  Roughly speaking, one defines a $\Z_2$-valued
function $\Om(x)=(-1)^{h(x)}$ on $H^3(V;\Z)$, obeying the usual relation
\eqn\uttu{\Om(x+y)=\Om(x)\Om(y)(-1)^{(x,y)}.}
This enables one to construct a theta function that determines the
partition function of the chiral two-form.\foot{This description omits
a twist that we recall in section 5.2.}

In general, there is no elementary formula for  $\Om(x)$.  However, 
for the case that the $M5$-brane can be related to a $D4$-brane,
there is such a formula, in part.  This is the case that $V=S\times R$, with
$S$ a circle with supersymmetric spin structure and $R$ a five-manifold.
In this case, we will justify the following assertion about $\Om(x)$:
if  $x$ is an element of $H^3(R;\Z)$, then
\eqn\troggo{h(x)=\int_R w_2(R)\cup x.}
Here to make sense of this integral, $x$ should be reduced mod 2,
and the integral is understood as an intersection number in mod 2
cohomology.  
To fully determine $\Om(x)$ (with the help of \uttu), we would also
need to compute $\Om(a\cup w)$
for $a$ a generator of $H^1(S;\Z)$
and $w\in H^2(R;\Z)$.  It does not seem that there is a formula
for $\Om(a\cup w)$ as elementary as \troggo.

In general, the physical application of $\Om(x)$ is rather subtle.
But (as in the case we considered in section 4), the interpretation
of $\Om(x)$ is more straightforward when $V=S\times R$.  In this case,
the chiral two-form on $V$ reduces on $R$ to an ordinary two-form field $B_2$
with field strength $T_3=dB_2$ and characteristic class $x=[T_3/2\pi]$,
or (by duality) to a one-form field $B_1$ with two-form
field strength $T_2=dB_1$
and characteristic class $v=[T_2/2\pi]$.  In the description by
a two-form field, the evaluation of the path integral includes a summation
over $x$ in which one must include the sign factor $\Om(x)$.  
This factor can be understood as coming from a term in the Lagrangian
\eqn\ikok{i\pi \int_R w_2\cup x.}

In the dual
description by a one-form field, the evaluation of the path integral
includes a summation over $v$.  In evaluating this sum, one includes
a sign factor $\Om(a\cup v)$ for which we will not obtain an explicit
general formula.  In addition
(as in the case considered in section 4), the interaction \ikok\ 
in the two-form  description is dual in the one-form description to a shift
in the periods of $T_2$.  The dual of \ikok\ is a shifted quantization law,
\eqn\ipolop{\int_U{T_2\over 2\pi}={1\over 2}\int_U w_2~{\rm mod}~\Z.}
The shift means that $B_1$, whose curvature is $T_2$,
is not a ``$U(1)$ gauge field,'' but
rather defines a $\spinc$ structure on $R$.  (Reciprocally, the sign
factor $\Om(a\cup v)$ will in general determine a shift in the periods
of $T_3$.)

Since $R$ might not be $\spinc$, something
is missing in the discussion so far.  There is an important
difference between \troggo\ and the analogous formula $h(x)=\int_R
\lambda\cup x$ that we met in section 4.  As $\lambda$ is an integral
cohomology class, the integral $\int_R\lambda\cup x$ vanishes
if $x$ is torsion; that is why torsion was not very important in
section 4.  However, $w_2$ is a $\Z_2$-valued cohomology class,
and $\int w_2\cup x$ can perfectly well be non-zero for torsion $x$.
We will show momentarily that precisely when $R$ is not $\spinc$,
there is a torsion class $x_0$ with $\Om(x_0)=-1$.  It follows
(since $(x,x_0)=0$ for all $x$, given that $x_0$ is torsion)
that $\Om(x+x_0)=-\Om(x)$ for all $x$.  In determining the partition
function of the $M5$-brane,  the factor $\Om(x)$ is the only factor
that is not invariant under $x\to x+x_0$.  (For example, since $x_0$ is
torsion, the ordinary kinetic energy of the two-form field does not
receive a contribution from $x_0$.)    The contributions to the partition
function from $x$ and $x+x_0$ will therefore cancel in pairs, and
the partition function of the $M5$-brane vanishes.  This vanishing
cannot be lifted by inserting local operators (which do not detect
a flat two-form field with characteristic class $x_0$), and so should
be understood as a sort of global anomaly.  Existence of this
anomaly gives an $M5$-brane
explanation of the fact that in Type IIA, the $D4$-brane world-volume
should be $\spinc$.

The existence of $x_0$ when $R$ is not $\spinc$ follows from some basic facts 
in algebraic topology.  The cup product gives a map
\eqn\turmigo{H^2(R;U(1))\times H^3(R;\Z)\to H^5(R;U(1)) = U(1)}
which by Poincar\'e and Pontryagin duality is a perfect pairing.
The ``perfectness'' means that every homomorphism $H^3(R;\Z)\to U(1)$
is $x\to \int_R \theta\cup x$ for some $\theta\in H^2(R;U(1))$,
and every homomorphism $H^2(R;U(1))\to U(1)$ is $\theta\to \int_R \theta
\cup x$ for some $x\in H^3(R;\Z)$.  If one restricts the pairing
in \turmigo\ to the torsion subgroup $H^3_{tors}(R;U(1))$, then one
gets an analogous perfect pairing
\eqn\urmigo{\bar H^2(R;U(1))\times H^3_{tors}(R;\Z)\to U(1).}
Here $\bar H^2(R;U(1))$ is the group of components of $H^2(R;U(1))$
(in other words, it is the quotient of $H^2(R;U(1))$ by the connected
component containing the identity).   The formula $h=\int_R w_2\cup x_0$
is equivalent to $\Om=\int_Ri(w_2)\cup x_0$ where $i:\Z_2
\to U(1) $ is the embedding of $\Z_2$ into $U(1)$.  So perfectness of
\urmigo\ means that a torsion class $x_0$ with $\Om(x_0)=-1$ exists precisely
if $i(w_2)$ is not in the identity component of $H^2(R;U(1))$.  Now
consider the commutative diagram
\eqn\kkoo{\matrix{ 0 &\to & \Z & \underarrow{2} &\Z & \to & \Z_2 & \to & 0 \cr
                      &    &\downarrow & & ~\downarrow \half
&& \downarrow i && \cr 
                   0 & \to &\Z &  \to & \R & \to & U(1) & \to & 0 \cr}}
where the first horizontal map in the top row
is multiplication by 2, the other horizontal maps are obvious inclusions
and reductions,
the first vertical map is the identity, the second vertical line is
multiplication by $1/2$, and the last is $i$.  Let $\beta:H^2(R;\Z_2)
\to H^3(R;\Z)$ be the Bockstein derived from the first row, and let
$\beta':H^2(R;U(1))\to H^3(R;\Z)$ be the Bockstein derived from the second.
The condition that $R$ is not $\spinc$ is $\beta(w_2)\not= 0$;
in fact, $W_3(R)=\beta (w_2)$ is the obstruction to $\spinc$ structure.
The condition that $i(w_2)$ not be in the identity component
of $H^2(R;U(1))$ is that $\beta'(i(w_2))\not= 0 $.  
Commutativity of the above diagram implies that $\beta'=i\beta$.
So $i(w_2) $ is not in the identity component, and a torsion $x_0$
with $\Om(x_0)=-1$ exists, if and only if $W_3(R)\not= 0$ and $R$ is
not $\spinc$.

\bigskip\noindent{\it Generalizations}

This discussion of a global anomaly is not limited to the case
that $V=S\times R$.  More generally, the $M5$-brane is anomalous
whenever there is a torsion class $x_0$ with $\Om(x_0)=-1$.
However, it is hard in general to give a criterion for existence of $x_0$.

I will now  briefly suggest how these anomalies can be removed by
turning on background fields.  
In the discussion so far, we have taken the Neveu-Schwarz
three-form field $H$ of Type IIA, and the corresponding $M$-theory
four-form field $G$, to be topologically trivial.  Naively, the classical
equations $dT_2=H $ and $dT=G$ (where $T_2$ is the two-form on
a $D4$-brane and $T$ is the self-dual three-form on an $M5$-brane)
imply that $H$ and $G$ should be trivial when restricted to
the $D4$- and $M5$-brane world-volumes.  However, taking into
account the global anomalies, the general statement for Type IIA is 
\refs{\branecft,\freed}%
\eqn\kko{H|_R = W_3(R),}
where $H|_R$ is shorthand for the restriction to $R$ of the characteristic
class of $H$.  
The analog of this condition
 for the $M5$-brane should apparently be the following.
Under the perfect pairing
\eqn\ujju{\bar H^3(V;U(1))\times H^3_{tors}(V;\Z)\to U(1)}
analogous to the one considered above,
the function $x_0\to \Om(x_0)$ (for $x_0$ torsion)
corresponds to an element $\theta\in \bar H^3(V;U(1))$.
The general statement about the restriction of $G$ to $V$ should
apparently be
\eqn\mko{G|_V=\beta'(\theta),}
where as above $\beta'$ is the Bockstein.  This reduces to \kko\ in the
appropriate situation, and I suspect that it holds in general.

\subsec{Direct Computation}

Let us next attempt to directly imitate the computation in
section 4.   To begin with, we assume that $V$ is spin.

For $x\in H^3(V;\Z)$, we want to define a suitable $\Z_2$-valued
function $\Om(x)=(-1)^{h(x)}$.  
We let $Z=S'\times V$ (with $S'$ a circle) and
set $z=a' \cup x$ with $a'$ a generator of $H^1(S';\Z)$.\foot{The
following computation has a very similar structure to the one
in section 4, although a few details are different.  To try  to bring
out the analogy, and hopefully without causing confusion,
we will use some of the notation of section
4 for objects that play the analogous role here.
The seven-manifold $Z$ is analogous to the eleven-manifold called
$Z$ in section 4; likewise, the eight-manifold $W$ of boundary $Z$ will
be analogous to the twelve-manifold called $W$ in section 4.
 Similarly, we will use the names
$S',a',x$, and $z$ for objects that play an analogous role to objects
of the same name in section 4.}
Then, assuming that $Z$ is the boundary of an eight-dimensional
spin manifold $W$ over which $z$ extends, one is tempted to set 
$h(x)=\int_W z\cup z$. 
This is not well-defined modulo 2, because in general for a closed
eight-dimensional spin manifold $W$, $\int_W z\cup z$  is not
even.  The quantity which is always even for a closed eight-dimensional
spin manifold with a given $z\in H^4(W;\Z)$ 
is $\int_W(z\cup z+\lambda\cup z)$ (where $\lambda$ is
the integral characteristic class with $2\lambda=p_1(W)$), so we set
\eqn\jilop{h(x)=\int_W(z\cup z+\lambda\cup z).}

Here we need, as in the analogous discussion
in section 4, to make sense of the integral $\int_W\lambda\cup z$
on the manifold-with-boundary $W$.
This integral needs some explanation, because in general
neither $\lambda$ nor $z$ vanishes on the boundary of $W$.
The approach taken in \witten\ was as follows.  If \jilop\ were
well-defined purely topologically, we would use the function $\Om(x)$
to quantize the torus $\T=H^3(V;\R)/H^3(V;\Z)$ that parametrizes
flat three-form fields $C$ on $V$ mod gauge transformations.
The $\lambda\cup z$ term in \jilop\ means that the torus that
we can naturally quantize is not $\T$ but the torus $\T'$ that
parametrizes, up to gauge transformations, $C$-fields of curvature
$\lambda/2$.  ($\T$ is isomorphic to $\T'$, by the map $C\to C+C_0$
where $C_0$ is any $C$-field of curvature $\lambda/2$, but there is
no canonical isomorphism between $\T$ and $\T'$.)  A heuristic
way to explain the shift from $\T$ to $\T'$
is that $z\to z-\lambda/2$ eliminates the $z\cup \lambda$ term in \jilop; for more information, see \witten.
An alternative approach to understanding the integral in \jilop\
 (described to me by M. Hopkins and I. M. Singer) is as follows.
The $\lambda$ class of a seven-dimensional
spin manifold such as $Z$ is always even.\foot{The intersection
form of the eight-manifold $B=\S^1\times V$ is even, so the
relation $\int_B(x\cup x +\lambda\cup x)\cong 0$ modulo 2 for
all $x\in H^4(B;\Z)$ reduces to $\int_B\lambda\cup x\cong 0$ modulo 2
for all $x$.  This implies that $\lambda$ is divisible by 2.}
Since we only want to define $h(x)$ modulo 2, we can interpret
the integral $\int_W\lambda\cup z$ as an integral in mod 2 cohomology,
replacing $\lambda$ and $z$ by their mod 2 reductions $\bar\lambda$
and $\bar z$.  Since $\bar\lambda$ vanishes when restricted to the
boundary of $W$, we can pick a trivialization of it; once such
a trivialization is picked, the integral $\int_W\bar\lambda \cup z$ makes
sense.  The relation between the two approaches is that a trivialization of
$\lambda$ mod 2  gives a way of identifying $\T$ and $\T'$.

The details in the last paragraph will not play a major role in the
present paper.  The reason is that, with $V=S\times R$, we will
compute $\Om(x)$ only for $x\in H^3(R;\Z)$.  This means that on
$Z=S'\times V=S'\times S\times R$, both $\lambda$ and $z=a'\cup x$
are pullbacks from $S'\times R$.  In trivializing $\lambda$ mod 2 on $Z$,
we can restrict ourselves to consider only
trivializations that are pulled back from $R$,
and the choice of such a trivialization does not affect the integral
$\int_W \lambda\cup z$.  At the level of differential forms, this last
statement means that under $\lambda\to\lambda+d\gamma$, $\int_W\lambda\cup z$
changes by $\int_{S'\times S\times R}\gamma \cup z$, which vanishes
for $\gamma$ and $z$ both being pullbacks from $S'\times R$.
Hence there is a completely canonical $\Om(x)$ for the $x$ we will consider,
and this is what we will evaluate.

Just as in section 4.2, it is inconvenient to calculate with $W$
required to be a spin manifold.  We can readily generalize the
discussion to permit $V$ and $W$ to be $\spinc$ manifolds, not necessarily
spin, as follows.  A $\spinc$ manifold $W$ (with a chosen $\spinc$ structure)
has a two-dimensional class $\alpha\in H^2(W;\Z)$, which reduces
mod 2 to $w_2(W)$.  In addition, on such a manifold $p_1-\alpha^2$
is divisible by 2, and there is an integral characteristic
class $\lambda$ such that\foot{More generally,
any real oriented vector bundle $E$ with $w_2(E)=0$ has an integral
characteristic class $\lambda$ with $2\lambda(E)=p_1(E)$.  If $W$ is
$\spinc$, let $J$ be a real two-plane bundle with Euler class $\alpha$,
and let $E=TW\oplus J$ (with $TW$ being the tangent bundle to $W$).  Then
$w_2(E)=0$, and $\lambda(E)$ is the desired class with
$2\lambda=p_1(E)=p_1(TW)-\alpha^2$.}  $2\lambda=p_1-\alpha^2$.
 Moreover, for any $x\in H^4(W;\Z)$,
$\int_W(x\cup x+\lambda\cup x)$ is always even.\foot{This can be
proved by generalizing the proof given in section 4 of \ugwitten\
(see eqn. (4.7)), where  $W$ was assumed to be spin.  
Let $J$ be a real two-plane bundle over $W$ with Euler class $\alpha$,
and let $N$ be the direct sum of $J$ with a trivial rank three bundle.
Let $K$ be a twelve-manifold that is the unit sphere bundle
in $N$; $K$ is spin.  Let $\pi:K\to W$ be the projection, let $x$
be any element of $H^4(W;\Z)$, and let $u$ be an element of $H^4(K;\Z)$
with $\pi_*(u)=1$ and $u\cup u=0$.  (Such a $u$ can be constructed
as the Poincar\'e dual of a section of $\pi$.)  Consider, as in
\ugwitten, an $E_8$ bundle $B$ over $K$ with characteristic class 
$u+\pi^*(x)$.
If $i(B)$ is the index of the Dirac operator on $K$ with values in $B$
(in the adjoint representation), then $i(B)$ is even (because $B$ is real
and $K$ has dimension of the form $8k+4$).  Evaluation
of $i(B)$ via the index theorem leads, as in \ugwitten\ (and using the 
fact that $\lambda(K)=\pi^*(\lambda(W))$ where $\lambda(W)$ is defined
as in the last footnote using the $\spinc$ structure of $W$),
to $i(B)=\int_W(x\cup x +\lambda(W) \cup x)$, and so this expression is even.}
So we can evaluate \jilop\ for any $\spinc$ manifold $W$, with $\lambda$
as just defined.

We will now consider
 $h(x)$ for $V=S\times R$.  We assume first that
$R$ is $\spinc$.  We give $V$
a $\spinc$ structure that is the product of the
supersymmetric (or unbounding) spin structure on $S$ with the given
$\spinc$ structure on $R$. We set $Z=S'\times V=S'\times S\times R$.
Suppose that $x\in H^3(R;\Z)$.  Then as in section 4.2, $Z$ is the
boundary of a $\spinc$ manifold $W=S'\times D\times R$, where $D$ is
a disc of boundary $S$; and $z=a'\cup x$ extends over $W$ as a pullback from
$S'\times R$.  The $\spinc$ structure on $W$ is the product
of a spin structure on $S'$, the given $\spinc$ structure on $R$ with
two-dimensional class $\alpha_R$, and a $\spinc$ structure on $D$ with
a two-dimensional class $\alpha_D$ such that $\int_D\alpha_D=1$.
(The reason for the last statement is the same as in section 4.2: the
supersymmetric spin structure on $S$ does not extend over $D$ as a spin
structure, but it extends as a $\spinc$ structure with $\int_D\alpha_D=1$.)
We have $p_1(W)=p_1(R)$ and
$\alpha(W)=\alpha_D+\alpha_R$; also, $\alpha_D\cup
\alpha_D=0$ since $D$ is two-dimensional.
We can compute the $\lambda$ class of $W$:
$\lambda(W)=(p_1(W)-\alpha(W)^2)/2=\lambda(R)-\alpha_D\cup\alpha_R$.
It follows that
\eqn\kliko{h(x)=\int_W\left(z\cup z +\lambda(w)\cup z\right)
=\int_W\left(z\cup z +\lambda(R)\cup z -\alpha_D\cup\alpha_R\cup z\right).}
On the right hand side, only the term $\alpha_D\cup\alpha_R\cup z$
contributes to the integral, as the others are pullbacks from $S'\times R$.
Using $z=a'\cup x$, with $\int_{S'}a'=1$, and $\int_D\alpha_D=1$, we
get
\eqn\rklik{h(x)=-\int_R \alpha_R\cup x.}
Since $\alpha_R$ is congruent to $w_2(R)$ mod 2, this is equivalent
to the promised formula \troggo.

So far we have assumed that $V$ is $\spinc$.
Otherwise,  the $\lambda$ class is no longer available, but we still
have the Wu class $v$ in mod 2 cohomology, with $\int_W(x\cup x +v \cup x)$
even.  In eight dimensions,
\eqn\jilko{v=w_2^2+w_4.}
So the definition of $h(x)$ should be
\eqn\koolo{h(x)=\int_W\left(z\cup z +(w_4+w_2^2)\cup z\right).}
Here we have given the most natural topological definition.
In section 5.3, we will verify that it is equivalent to the physics-based definition in \witten.

In the meantime, we can use \koolo\ to show that \troggo\ is true
for all $V=S\times R$ and $x\in H^3(R;\Z)$, whether or not $R$ is
$\spinc$.  For this, we note that it follows from \jilko\ that if $R$
is the boundary of an oriented manifold $\tilde R$ over which $x$ extends,
then $h(x)=0$.  For in this case, we can set $W=S'\times S\times \tilde R$,
and the integral defining $h(x)$ vanishes as $x,$ $w_2$, and $w_4$ are
all pullbacks from $S'\times \tilde R$.  This bordism property can be used
to reduce to the case that $R$ is $\spinc$.  Indeed, we can always find
an oriented six-manifold  
$\tilde R$ whose boundary is $R-R_1-R_2$ (the minus signs keep track of the
orientations), where $x$ extends over $\tilde R$ and vanishes on $R_2$, and $R_1$
is $\spinc$.\foot{The precise mathematical statement here is that
$\Omega_5(K(\Z,3))$, the bordism group of oriented five-manifolds
equipped with a three-dimensional cohomology class $x$, is $\Z_2\times \Z_2$,
a complete set of
 invariants being $\int w_2\cup x$ and $\int w_2\cup w_3$.  
(This statement and analogous ones cited in the next paragraph
were provided by R. Stong,
along with   proofs.)
So for the bordism
group, we can pick two generators $R_1'$ and $R_2'$, where 
$\int w_2\cup x$ is nonzero on $R_1'$ and zero on $R_2'$,
and $\int w_2\cup w_3$ is nonzero on $R_2'$ and zero on $R_1'$.
Moreover, one can pick $R_1'$ to be $\spinc$, and one can assume that
$x$ vanishes on $R_2'$.  The fact that $R_1'$ and $R_2'$ generate
the bordism group means that $R-R_1-R_2$ is a boundary, where
the $R_i$ are as in the text and each $R_i$ is equal to $R_i'$ or empty,
depending on the values of the invariants of $R$.}
The bordism property implies that $h(x)$ is the same
whether computed on $S\times R$ or $S\times R_1$ ($R_2$ does not
contribute as $x$ vanishes on $R_2$).  As $R_1$ is $\spinc$,
we can use our previous result: $h(x)=\int_{R_1}w_2\cup x$.  Since the
characteristic class $w_2(R)$ automatically
extends over $\tilde R$, one has $\int_Rw_2\cup x=\int_{R_1}w_2\cup x$.
Hence $h(x)=\int_R w_2\cup x$ whether or not $R$ is $\spinc$.

We could have made a much more extensive use of bordism
in the present paper.  Indeed, we could have used the fact  that $\Omega_5^{\spinc}(K({\bf Z},2))=\Z$,
generated by $\int \alpha\cup x$, to show that \troggo\ is the only
 nonzero bordism-invariant formula for $h(x)$ in the $\spinc$ case, whereupon
we could deduce from the example considered in section 2 that \troggo\
is correct.  We similarly
could use the fact that $\tilde \Omega_9^{{\rm Spin}}(K({\bf Z},5))=\Z$,
generated by $\int\lambda\cup x$, plus invariance under bordism,
to reduce the computation in section
4.2 to the special case considered in section 2.  This would give
short cuts to the desired results, but we have chosen instead to base
our computations on a better understanding of the formalism in \witten.

\subsec{Comparison To Physical Definition}

It remains to compare the obvious topological definition \koolo\ to
the physics-based formalism in \witten.
The full physical setup for this problem
depends on details that we have so far omitted.  
The $M5$-brane worldvolume $V$ is embedded in an eleven-manifold
$M$.  $V$ is orientable (but not necessarily spin), and $M$ is spin.
Let $N$ be the normal bundle to $V$ in $M$.  The condition for $M$ to
be spin is
\eqn\polpo{w_1(N)=0,~~ w_2(N)=w_2(V).}
Also, the Euler class of $N$ vanishes (or equivalently, as $N$ is of
odd rank, $w_5(N)=0$), for reasons explained in section 5 of \ugwitten.
Another part of the data is the four-form field $G$ of $M$-theory.
It is of the form
\eqn\nolpo{{G\over 2\pi}={\lambda(M)\over 2}+g,}
where $g$ is an integral class.  Moreover, if $U$ is a small four-sphere
linking $V$ in $M$, then 
\eqn\mikol{\int_Ug=1,}
 since the fivebrane has unit charge.

Let $P$ be the submanifold of $M$ consisting of all points a distance
$\epsilon$ from $V$, for some very small $\epsilon$.  $P$ is a  four-sphere
bundle over $V$.  Let $\pi:P\to V$ be the projection.  \mikol\ is
the statement that \eqn\kloo{\pi_*(g)=1.}
This uniquely determines $g$ modulo $g\to g+\pi^*(y)$ for $y\in H^4(V;\Z)$.
Note that $\pi_*(g\cup g)$ is invariant mod 2 under such a transformation
of $g$.  Hence, its mod 2 reduction does not depend on the choice of $g$.
In fact,
\eqn\jurry{\pi_*(g\cup g)\cong w_4(N)~{\rm mod}~2.}
To prove this, since the left hand side is independent of the choice
of $g$ modulo 2, it suffices to consider the case that $g$ is the Poincar\'e
dual to a section $s$ of $\pi$.  (Such a section exists at least
over the five-skeleton of $V$, since the Euler class of $N$ is zero, 
and a choice of $s$ on the five-skeleton suffices
for evaluating the four-dimensional class on the left hand side of \jurry.)
Choice of such a section splits $N$ as $N={\cal O}\oplus N'$ where
${\cal O}$ is a rank one trivial real bundle (consisting of multiples of $s$)
and $N'$ is a rank four bundle.  $g\cup g$ is Poincar\'e dual to the
intersection class $s\cap s$.  If we regard $s$ as a codimension
four submanifold of $P$, then its normal bundle is $N'$, so
$s\cap s$ is dual to the restriction to $s$ of the Euler class
$\chi(N')$, and hence $\pi_*(g\cup g)=\pi_*(g\cup \chi(N'))=\pi_*(g)\cup
\chi(N')=\chi(N')$.  But (for any $SO(4)$
bundle $N'$) $\chi(N')$ is
congruent to $w_4(N')$ mod 2, and with $N={\cal O}\oplus N'$, we have
$w_4(N)=w_4(N')$.  This justifies the assertion in \jurry.

Pick a class $x\in H^3(V;\Z)$.  We will now restate the definition
of $\Om(x)=(-1)^{h(x)}$ given in \witten. 
Let $z=a'\cup x\in H^4(S'\times V;\Z)$, with $a'$ a generator of $H^1(S';\Z)$. 
Let $\tilde 
Z=S'\times P$, where $S'$ is a circle with Neveu-Schwarz spin structure.
Thus, $\tilde Z$ is a four-sphere bundle over $Z=S'\times V$; we write
$\tilde \pi$ for the projection $\tilde\pi:\tilde Z\to Z$.
And define $w\in H^4(\tilde Z;\Z)$ by  $w=\tilde \pi^*(z)+g$.  Let now $\tilde
W$ be a twelve-dimensional spin manifold
with boundary $\tilde Z$ over which $w$ extends.   Such a $\tilde W$ always exists 
\ref\stong{R. Stong, ``Calculation Of $\Omega_{11}^{spin}(K({\bf Z},4)$,''
in {\it Unified String Theories}, eds. M. B. Green and D. J. Gross
(World Scientific, 1986).}.
The definition in \witten\ can be stated
\eqn\olop{h(x)={1\over 3}\int_{\tilde W}\left(w-\half \lambda\right)\left(
(w-\half\lambda)^2-{1\over 8}(p_2-\lambda^2)\right)-\left(w\to 0\right).}
Here the meaning of the last term is that one should subtract the same
expression with $w$ replaced by 0.  $E_8$ index theory is used to prove
that $h(x)$ is integral and independent modulo 2
of the choice of $\tilde W$ and of the extension of $w$.  The fact that the class
that is integrated in \olop\ is not canonically trivial near the boundary
means that the function $\Om(x)$ enables us to quantize not the space
$H^3(V;U(1))$ of flat three-form fields on $V$, but a shifted version of it.

The definition of $h(x)$ just given is rather abstract.  For computation,
it is convenient to make some simplifying assumptions that are actually
rather mild in practice.  Suppose that $S'\times V$ is the boundary
of an oriented eight-manifold $W$ over which $N$ extends (as a rank
five bundle obeying $w_1(N)=0$, $w_2(N)=w_2(W)$, and $w_5(N)=0$).
Let $\tilde W$ be the unit sphere bundle in $N$; the conditions on $N$ ensure
that $\tilde W$ is spin, and its boundary is $\tilde Z=S'\times P$.  Suppose
further that $z=a'\cup x$ extends over $W$, and  that $g$
extends over $\tilde W$.  Then, setting $w=\pi^*(z)+g$, we can simplify
\olop\ by integrating over the fibers of $\pi:\tilde W\to W$.  We get
\eqn\nikkop{h(x)=\int_W\left(z \cup z -z\cup \lambda(\tilde W)
+z\cup \pi_*(g\cup g)\right).}
(We have dropped terms that vanish if $x=0$; they in fact vanish mod
2 using the fact that the integral in \olop\ is even if evaluated
on a closed twelve-dimensional spin manifold, and the fact that 
the integrand vanishes near the boundary if $x=0$.)

To clarify this further, we would like to express the mod 2 reduction
of $-\lambda(\tilde W)+\pi_*(g\cup g)$ in terms of quantities defined just
on $W$.  For this, we note first that (for any spin manifold
$\tilde W$) $\lambda(\tilde W)$ is congruent
mod 2 to $w_4(\tilde W)$.  Stably, the tangent bundles $T\tilde W$ and $TW$ of 
$\tilde W$ and 
$W$ are related by  $T\tilde W=TW\oplus N$.
So since $w_1(W)=w_1(N)=0$ and $w_2(N)=w_2(W)$, we have
$w_4(\tilde W)=w_4(W)+w_2(W)w_2(N)+w_4(N)=w_4(W)+w_2(W)^2+w_4(N)$.  
Using also \jurry, we learn that
$-\lambda(\tilde W)+\pi_*(g\cup g)$ 
is congruent mod 2 to $w_4(W)+w_2(W)^2$, so
that \nikkop\ is equivalent to
\eqn\hikkopo{h(x)=\int_W\left(z\cup z+w_4(W)\cup z +w_2(W)\cup w_2(W)\cup z\right).}
This is the formula that we guessed
on purely formal grounds toward the end of section 5.2.
What we have gained is an understanding of how this
formula is related to eleven-dimensional physics.

\bigskip
This work was supported in part by NSF Grant PHY-9513835 and the
Caltech Discovery Fund.  I am grateful to M. J. Hopkins and I. M. Singer
for numerous
explanations of their viewpoint about the fivebrane action, as well
as other matters.  In addition, I would like to thank G. Moore for 
discussions and suggestions about the manuscript, R. Stong for
helpful correspondence, and E. Diaconescu, D. Freed,
and A. Kapustin for comments and questions.
\listrefs
\end